\documentclass[twocolumn,showpacs,preprintnumbers,amsmath,amssymb,prl,
superscriptaddress]{revtex4}
\usepackage{graphicx}
\usepackage{dcolumn}
\usepackage{bm}
\usepackage{amsmath}
\usepackage{hyperref}
\usepackage[dvips]{color}

\def\rnum#1{\expandafter{%
\romannumeral #1}}
\def\Rnum#1{\uppercase\expandafter{%
\romannumeral #1}}

\newcommand{\bol}[1]{\boldsymbol #1}

\begin{document}

\title{
Spin-Nematic and Spin-Density-Wave Orders in Spatially Anisotropic
Frustrated Magnets in a Magnetic Field
}

\author{Masahiro Sato}
\affiliation{Department of Physics and Mathematics, Aoyama Gakuin University,
Sagamihara, Kanagawa 229-8558, Japan}
\author{Toshiya Hikihara}
\affiliation{Faculty of Engineering, Gunma University, Kiryu, Gunma 376-8515,
Japan}
\author{Tsutomu Momoi}
\affiliation{Condensed Matter Theory Laboratory, RIKEN, Wako, Saitama
351-0198, Japan}
\date{\today}

\begin{abstract}
We develop a microscopic theory of finite-temperature spin-nematic orderings 
in three-dimensional spatially anisotropic magnets consisting of
weakly-coupled frustrated spin-$\frac{1}{2}$ chains
with 
nearest-neighbor and 
next-nearest-neighbor couplings in a magnetic field.
Combining a field theoretical technique with 
density-matrix renormalization group results,
we complete finite-temperature phase diagrams
in a wide magnetic-field range that possess spin-bond-nematic and
incommensurate spin-density-wave ordered phases. The effects of a
four-spin interaction are also studied. The relevance of our results to
quasi-one-dimensional edge-shared cuprate magnets
such as LiCuVO$_4$ is discussed.
\end{abstract}


\maketitle

\textit{Introduction}.$-$
The quest for novel states of matter has been attracting much attention
in condensed-matter physics.
Among those states, recently spin-nematic (quadrupolar)
phases have been vividly discussed in the field of frustrated 
magnetism~\cite{Momoi1,Vekua,Hikihara1,Sudan,Sato09,Sato11,
Penc_Lauchli,Shindou,Zhitomirsky,Nishimoto}.
The spin-nematic phase is defined 
by the presence of
a symmetrized rank-2 spin tensor order, such as
$\langle S^+_{\bol r}S^+_{\bol r'}+{\rm H.c.}\rangle\ne0$, and
the absence of any spin (dipolar) moment. 
Geometrical frustration, which generally suppresses spin orders,
is an important
ingredient for the emergence of spin nematics~\cite{Momoi1}.
In spin-$\frac{1}2$ magnets, the spin nematic operators cannot be defined
on a single site because of the commutation relation of
spin-$\frac{1}{2}$ operators. They reside on {\it bonds} between
different sites~\cite{Momoi1,Hikihara1}, which is a significant difference
from the quadrupolar phases in higher-spin systems~\cite{Penc_Lauchli}.
Due to this property, it is generally quite hard to develop
theories of spin nematics in spin-$\frac{1}2$ magnets, particularly
in two- or three-dimensional (3D) systems. 
Developing such a theory is a current important issue in magnetism.


Among the existing models
predicting spin-nematic phases,
the spin-$\frac{1}2$ frustrated chain with a ferromagnetic 
nearest-neighbor coupling $J_1<0$ and an antiferromagnetic (AF)
next-nearest-neighbor one $J_2>0$ would be
the most relevant in nature 
because this system is believed to be an 
effective model for a series of quasi-1D edge-shared cuprate magnets such as
$\rm LiCuVO_4$~\cite{Enderle,Hagiwara,Masuda_VO4,Mourigal,Svistov,Takigawa},
$\rm Rb_2Cu_2Mo_3O_{12}$~\cite{Hase},
$\rm PbCuSO_4(OH)_2$~\cite{Yasui,Wolter}, $\rm LiCuSbO_4$~\cite{Dutton},
and $\rm LiCu_2O_2$~\cite{Masuda2}. 
These quasi-1D magnets hence offer a 
promising playground for spin-nematic phases.

Low-energy properties of the spin-$\frac{1}2$ $J_1$-$J_2$ chain have
been well understood thanks to recent 
theoretical efforts~\cite{Vekua,Hikihara1,Sudan,Sato09,Sato11}.
The corresponding Hamiltonian is given by
\begin{eqnarray}
\label{eq:J1J2chain}
{\cal H}&=&\sum_{n=1,2}\sum_j J_n{\bol S}_j\cdot{\bol S}_{j+n}-H\sum_j S_j^z,
\end{eqnarray}
where ${\bol S}_j$ is the spin-$\frac{1}{2}$ operator on site $j$ and
$H$ is an external field.
Below the saturation field in the broad parameter range $-2.7\alt J_1/J_2<0$,
the nematic operator $S_j^\pm S_{j+1}^\pm$ and
the longitudinal spin $S_j^z$ exhibit quasi-long-range orders,
while the transverse spin correlator $\langle S_j^{\pm}S_0^\mp\rangle$
decays exponentially 
due to the formation of two-magnon
bound states~\cite{Hikihara1}. This phase is called a spin-nematic
Tomonaga-Luttinger (TL) liquid, and it expands down to a
low-field regime. The nematic correlation is stronger than
the incommensurate longitudinal spin correlation in the high-field regime,
while the latter grows stronger in the low-field regime.

From these theoretical results, 
the quasi-1D cuprates are
expected to possess incommensurate longitudinal spin-density-wave (SDW)
and spin-nematic long-range orders, respectively, in low- and high-field 
regimes at sufficiently low temperatures. In fact, recent magnetization
measurements of $\rm LiCuVO_4$ at
low temperatures have detected a new
phase~\cite{Hagiwara} near saturation,
and it is expected to be a 3D spin nematic phase. Some experiments
on $\rm LiCuVO_4$ in an intermediate-field regime find 
SDW oscillations~\cite{Masuda_VO4,Mourigal,Svistov}
whose wave vectors agree with 
the result of the nematic TL-liquid theory~\cite{Vekua,Hikihara1,Sato09}.
Furthermore, the spin dynamics of $\rm LiCuVO_4$ observed by
NMR~\cite{Takigawa} seems to be consistent with the prediction
from the same theory~\cite{Sato09,Sato11}. 
However, this nematic TL-liquid picture can be applicable 
only above the 3D ordering temperatures. 
We have to take into account interchain
interactions to explain how 3D spin-nematic and SDW long-range ordered
phases are induced with lowering temperature.
A mean-field theory for the 3D nematic phase of 
quasi-1D spin-$\frac{1}2$ magnets~\cite{Zhitomirsky} has
been proposed recently, but it cannot be applied to the SDW phase
and does not quantitatively describe finite-temperature effects. 
It is obscure how both nematic and SDW ordered phases are described
in a unified way. A reliable theory for 3D orderings
in weakly coupled spin-$\frac{1}2$ $J_1$-$J_2$ chains
is strongly called for.

In this Letter, 
we develop a general theory for
spin nematic and incommensurate SDW orders in spatially anisotropic
3D magnets consisting of weakly coupled $J_1$-$J_2$ spin chains with
arbitrary interchain couplings in a {\it wide magnetic-field range}.
Combining 
field theoretical and numerical results for
the $J_1$-$J_2$ spin chain,
we obtain {\it finite-temperature} phase diagrams,
which contain both spin-nematic and SDW phases
at sufficiently low temperatures.
We thereby reveal characteristic features in the ordering of
weakly coupled $J_1$-$J_2$ chains, which cannot be predicted from
the theory for the single $J_1$-$J_2$ chain. We also discuss the relevance
of our results to real compounds such as $\rm LiCuVO_4$.

\begin{figure}
\begin{center}
\includegraphics[width=7cm]{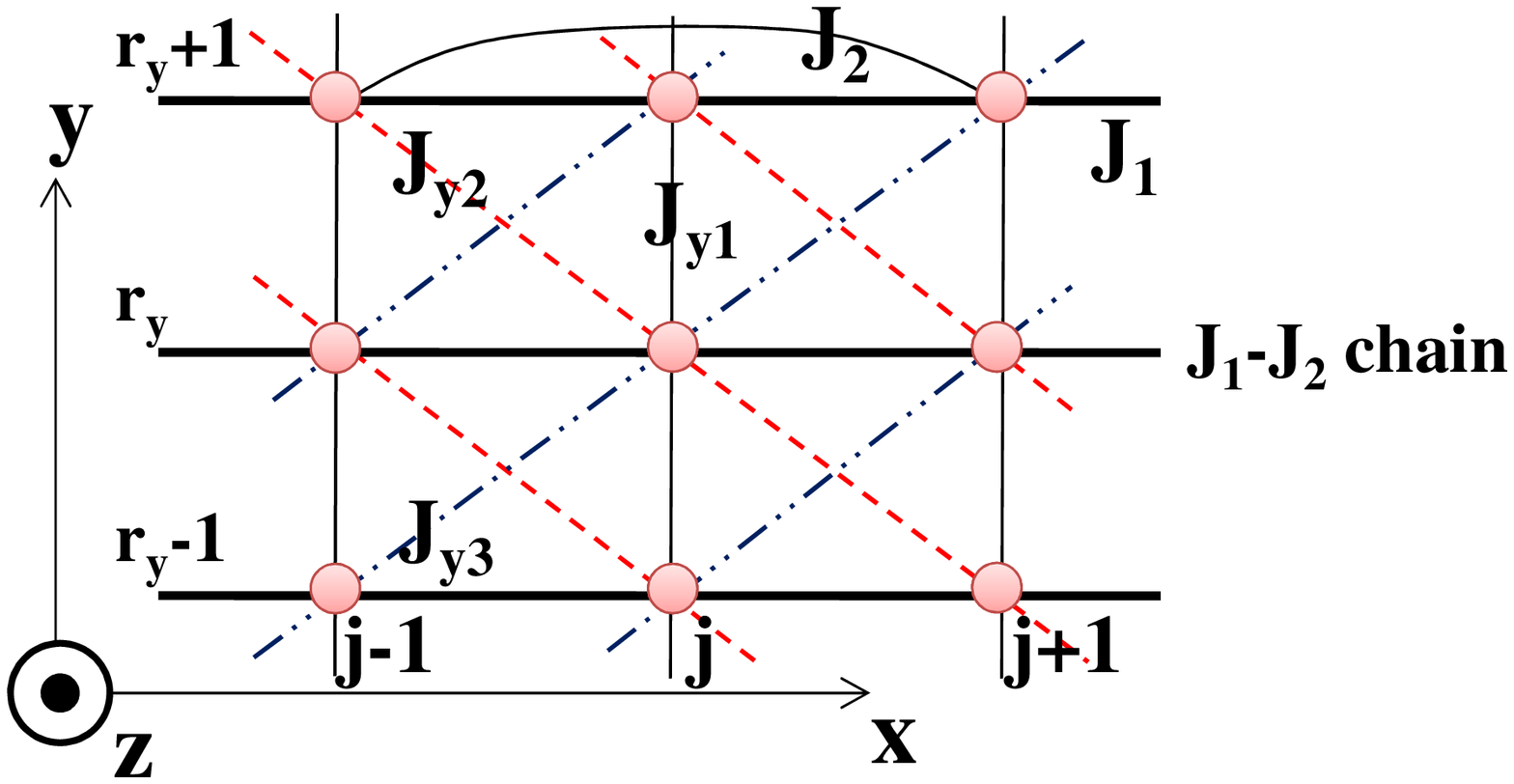}
\end{center}
\caption{(color online) Spatially anisotropic spin model consisting
of weakly coupled spin-$\frac{1}{2}$ $J_1$-$J_2$ chains.
We introduce inter-chain couplings $J_{y_1,y_2,y_3}$ 
in the $x$-$y$ plane. Similarly, 
$J_{z_1,z_2,z_3}$ are present in the $x$-$z$ plane. }
\label{fig:Model}
\end{figure}

\textit{Model}.$-$ 
Our model of a spatially anisotropic magnet is depicted 
in Fig.~\ref{fig:Model}. 
The corresponding Hamiltonian is expressed as
\begin{eqnarray}
\label{eq:Model}
{\cal H}_{3D}&=&\sum_{\bol r} {\cal H}_{\bol r}+{\cal H}_{\rm int},
\end{eqnarray}
where ${\bol r}=(r_y,r_z)$ denotes the site index of the square lattice
in the $y$-$z$ plane, ${\cal H}_{\bol r}$ denotes the
Hamiltonian~(\ref{eq:J1J2chain}) for the ${\bol r}$-th
$J_1$-$J_2$ chain along the $x$ axis in magnetic field $H$,
and ${\cal H}_{\rm int}$ is the inter-chain interaction.
In ${\cal H}_{\rm int}$, we introduce weak inter-chain Heisenberg-type
exchange interactions with coupling constants $J_{y_i}$ and
$J_{z_i}$ ($i=1,2,3$) defined in the $x$-$y$ and $x$-$z$ planes,
respectively~\cite{note1}.

\textit{Spin-$\frac{1}{2}$ $J_1$-$J_2$ chain}.$-$
Under the condition
$|J_{y_i,z_i}|\ll |J_{1,2}|$, it is reasonable to choose
decoupled $J_1$-$J_2$ spin chains (${\cal H}_{\bol r}$) as
the starting point for analyzing the 3D model (${\cal H}_{3D}$).
The low-energy effective Hamiltonian for the nematic TL-liquid
phase 
is given by
\begin{align}
\label{eq:Effective_J1J2chain}
{\cal H}_{\rm eff}^{\bol r}&= \int dx \sum_{\nu=\pm}\frac{v_\nu}{2}
\left[K_\nu(\partial_x\theta_\nu^{\bol r})^2
+K_\nu^{-1}(\partial_x\phi_\nu^{\bol r})^2\right]\nonumber\\
&+G_- 
\sin(\pi M)\sin(\sqrt{4\pi}\phi_-^{\bol r}+\pi M),
\end{align}
where $x=a_0 j$ (the length $a_0$ of the $J_1$ bond is set equal to unity),
$(\phi_\pm^{\bol r}(x),\theta_\pm^{\bol r}(x))$ is the canonical pair
of scalar boson fields, and $v_\pm$ and $K_\pm$ are, respectively,
the excitation velocity and the TL-liquid parameter of the
$(\phi_\pm,\theta_\pm)$ sector.
The sine term 
makes $\phi_-$ pinned, inducing an excitation gap
in the $(\phi_-,\theta_-)$ sector. Physically,
the gap corresponds to the magnon binding energy $E_b$.
On the other hand, the $(\phi_+,\theta_+)$ sector describes 
a massless TL liquid. Vertex operators are renormalized as
$\langle e^{i\alpha\sqrt{\pi}\phi_+(x)}
e^{-i\alpha\sqrt{\pi}\phi_+(0)}\rangle_+
=|2/x|^{\alpha^2K_+/2}$ for $|x|\gg 1$, in which
$\langle\cdots\rangle_\pm$ denotes the average over the
$(\phi_\pm,\theta_\pm)$ sector.
Spin operators ${\bol S}_{j,{\bol r}}$ are also bosonized as
\begin{subequations}
\label{eq:Spin}
\begin{align}
\label{eq:Spinz}
& S_{j,{\bol r}}^z \approx M+\partial_x
(\phi_+^{\bol r}+(-1)^j\phi_-^{\bol r})/\sqrt{\pi}
\nonumber\\
&+(-1)^q A_1 \cos[\sqrt{\pi}(\phi_+^{\bol r}+(-1)^j\phi_-^{\bol r})+2\pi Mq]
+\cdots,\\
& S_{j,{\bol r}}^+ \approx
e^{i\sqrt{\pi}(\theta_+^{\bol r}+(-1)^j\theta_-^{\bol r})}
\big\{(-1)^qB_0
\nonumber\\
& +B_1
\cos[\sqrt{\pi}(\phi_+^{\bol r}+(-1)^j\phi_-^{\bol r})+2\pi Mq]
+\cdots\big\},
\label{eq:Spin+-}
\end{align}
\end{subequations}
where $M=\langle S^z_{j,{\bol r}}\rangle$, $q=\frac{j}{2}$ ($\frac{j-1}{2}$) 
for even (odd) $j$, and $A_n$ and $B_n$ are nonuniversal constants.
Utilizing Eqs.~(\ref{eq:Effective_J1J2chain}) and (\ref{eq:Spin}), we can
evaluate spin and nematic correlation functions at zero temperature
($T=0$) as follows~\cite{Hikihara1,Sato09,Sato11}:
\begin{subequations}
\label{eq:correlators}
\begin{align}
\label{eq:correlators1}
& \langle S_j^+S_0^-\rangle
\approx  B_0^2 \cos(\pi j/2)
(2/|j|)^{1/(2K_+)}g_-(x)
+\cdots,
\\
& \langle S_j^zS_0^z\rangle
\approx  M^2 + (A_1^2/2)
|\langle e^{i\sqrt{\pi}\phi_-}\rangle_-|^2
\nonumber\\
&\hspace{1.7cm}\times\cos[\pi j(M-1/2)] (2/|j|)^{K_+/2}+\cdots,
\label{eq:correlators2}
\\
& \langle S_j^+S_{j+1}^+  S_0^-S_1^-\rangle
 \approx  (-1)^j 
C_0|j|^{-2/K_+}+\cdots,
\label{eq:correlators3}
\end{align}
\end{subequations}
where $g_-(x)=\langle e^{\pm i\sqrt{\pi}\theta_-(x)}
e^{\mp i\sqrt{\pi}\theta_-(0)}\rangle_-$,
$C_0$ is a constant and we have omitted the index ${\bol r}$.
The function $g_-(x)$ decays exponentially as $x^{-1/2}e^{-x/\xi_-}$.
The parameter $K_+$, which is less than 2 in the
low magnetization regime, monotonically increases with 
$M$~\cite{Hikihara1} and $K_+\to 4$ at the saturation.
Thus, the spin-nematic (SN) correlation is stronger than the incommensurate 
SDW correlation in the high-field regime with $K_+>2$ and weaker in the 
low-field regime with $K_+<2$. 

\begin{figure}
\begin{center}
\includegraphics[width=7.5cm]{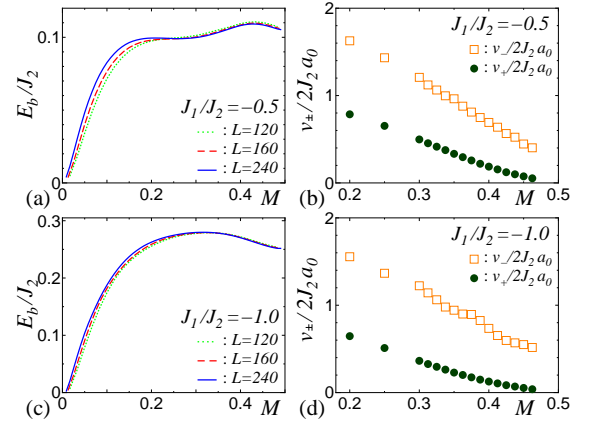}
\end{center}
\caption{(color online) (a),(c) Magnon binding energy $E_b$ 
and (b),(d) excitation velocities $v_\pm$ (b)(d)
as a function of $M$
in the spin-nematic TL-liquid phase in the spin-$\frac{1}{2}$
$J_1$-$J_2$ chain at $T=0$.
}
\label{fig:Velocity_Gap}
\end{figure}
The correlation length $\xi_-$ is related to $v_-$ via $v_-=\xi_- E_b$
under the assumption that the low-energy theory for the
$(\phi_-,\theta_-)$ sector has Lorentz invariance.
The velocity $v_+$ has the relation $v_+=2K_+/(\pi \chi)$, where
$\chi=\partial M/\partial H$ is the uniform susceptibility.
Since $K_+$, $\xi_-$, $E_b$, and $\chi$ are all determined 
with reasonable accuracy by using the density-matrix renormalization group
(DMRG) method~\cite{Hikihara1,H-F}, $v_\pm$ can be quantitatively evaluated 
as depicted in Fig.~\ref{fig:Velocity_Gap}.
The figure shows that $v_-$ is always larger than $v_+$, 
in accordance with the perturbative formulas 
$v_\pm\approx v(1\pm KJ_1/(\pi v)+\cdots)$ for $|J_1|\ll J_2$, 
in which $v$ and $K$ are
respectively the spinon velocity and the TL-liquid parameter for
the single AF-$J_2$ chain. We also note that $v_+$ approaches zero
at $M\to \frac{1}{2}$.

\textit{Analysis of the 3D model}.$-$ Let us now analyze the 3D model
(\ref{eq:Model}) starting with the effective theory of the $J_1$-$J_2$ chain.
We first bosonize all of the inter-chain couplings in ${\cal H}_{\rm int}$
through Eq.~(\ref{eq:Spin}).
To obtain the low-energy effective theory for
Eq.~(\ref{eq:Model}), we trace out the massive
$(\phi_-^{\bol r},\theta_-^{\bol r})$ sectors in the Euclidean action
${\cal S}_{\rm tot}={\cal S}_0+{\cal S}_{\rm int}$ via the cumulant expansion
${\cal S}_{\rm eff}^{3D}= {\cal S}_0+\langle {\cal S}_{\rm int}\rangle_-
-\frac{1}{2}\left(\langle {\cal S}_{\rm int}^2\rangle_-
-\langle {\cal S}_{\rm int}\rangle_-^2\right)
+\cdots$,
where ${\cal S}_0$ and ${\cal S}_{\rm int}$ are, respectively, the action
for the TL-liquid part of the $(\phi_+^{\bol r},\theta_+^{\bol r})$ sectors
and that for the inter-chain couplings.
This 
corresponds to the series expansion in $J_{y_i,z_i}/v_-$.
The resultant effective Hamiltonian is
expressed as ${\cal H}_{\rm eff}^{3D}={\cal H}_0+{\cal H}_{\rm SDW}
+{\cal H}_{\rm SN}+\cdots$. Here, ${\cal H}_0=\sum_{\bol r}\int dx
\frac{v_+}{2}\left[K_+(\partial_x\theta_+^{\bol r})^2
+K_+^{-1}(\partial_x\phi_+^{\bol r})^2\right]$ is the TL-liquid part 
and ${\cal H}_{\rm SDW}$ and ${\cal H}_{\rm SN}$ are, respectively,
obtained from the first- and second-order cumulants as follows:
\begin{subequations}
\label{eq:Effective_3d}
\begin{align}
{\cal H}_{\rm SDW} =
G_{\rm SDW}&\int \frac{dx}{2}
\sum_{{\bm r}}
\sum_{\alpha=y,z \atop ({\bm r}'={\bm r}+{\bm e}_\alpha)}
\left[J_{\alpha1}\cos(\sqrt{\pi}(\phi_+^{\bol r}-\phi_+^{{\bol r}'}))\right.
\nonumber\\
&
-J_{\alpha2}\sin(\sqrt{\pi}(\phi_+^{\bol r}-\phi_+^{{\bol r}'})-\pi M)
\nonumber\\
&
+\left. J_{\alpha3}\sin(\sqrt{\pi}(\phi_+^{\bol r}-\phi_+^{{\bol r}'})+\pi M)
\right],
\\
{\cal H}_{\rm SN} = 
G_{\rm SN}&
\int \frac{dx}{2}
\sum_{{\bm r}}
\sum_{\alpha=y,z \atop ({\bm r}'={\bm r}+{\bm e}_\alpha)}
\left[
{J_{\alpha1}}^2
-(J_{\alpha2}-J_{\alpha3})^2\right]\nonumber\\
&
\times \cos(\sqrt{4\pi}(\theta_+^{\bol r}-\theta_+^{{\bol r}'}))
\end{align}
\end{subequations}
with coupling constants
$G_{\rm SDW}=A_1^2|\langle e^{i\sqrt{\pi}\phi_-}\rangle_-|^2$~\cite{note2}
and $G_{\rm SN}=-\frac{B_0^4}{4v_- }\int dx v_-d\tau g_-(x,\tau)^2$
($\tau$ is imaginary time). The summations run over all nearest neighbor pairs
of chains,
where ${\bm r}^\prime={\bm r}+{\bm e}_\alpha$ ($\alpha=y,z$),
${\bm e}_\alpha$ denotes the unit vector along the $\alpha$-axis, and
we have assumed that the field $\phi_+$ smoothly varies in $x$.
The first-order term ${\cal H}_{\rm SDW}$ contains
an inter-chain interaction between
the operators $e^{\pm i\sqrt{\pi}\phi^{\bol r}_+}$,
which essentially induces a 3D spin longitudinal order.
Similarly,
the term ${\cal H}_{\rm SN}$ contains an inter-chain interaction
between the spin-nematic operators
$S_{j,{\bol r}}^\pm S_{j+1,{\bol r}}^\pm \sim (-1)^j
e^{\pm i\sqrt{4\pi}\theta_+^{\bol r}}$, which enhances a 3D spin nematic
correlation. We should notice that the effective
theory ${\cal H}_{\rm eff}^{3D}$ is reliable under the condition that
temperature $T$ is sufficiently smaller than
the binding energy $E_b$ and the velocities $v_\pm$.

Both the couplings $G_{\rm SDW,SN}$ can be numerically evaluated 
from the DMRG data of correlation functions~\cite{Hikihara1,H-F}:
$G_{\rm SDW}$ corresponds to the amplitude of the leading term of
the longitudinal correlator $\langle S_j^zS_0^z\rangle$ given
in Eq.~(\ref{eq:correlators}) and $G_{\rm SN}$ can be evaluated as
$G_{\rm SN}\approx \pi v_-^{-1}
\sum_{j=1}^{L}(j/2)^{1/K_+}j\langle S_j^+ S_0^-\rangle^2$.
We have checked that the finite-size correction to the sum
is small enough when the cutoff
$L$ is larger than $\xi_-$.
We emphasize that {\it there is no free parameter
in 
${\cal H}_{\rm eff}^{3D}$}.

To obtain the finite-temperature phase diagram, we apply the inter-chain
mean-field (ICMF) approximation~\cite{Scalapino,Bocquet} to the effective
Hamiltonian ${\cal H}_{\rm eff}^{3D}$.
To this end, we introduce the ``effective" SDW operator
${\cal O}_{\rm SDW}=e^{i\pi (\frac{1}{2}-M)j}e^{i\sqrt{\pi}\phi_+^{\bol r}}$
and the spin-nematic operator
${\cal O}_{\rm SN}=(-1)^j e^{i\sqrt{4\pi}\theta_+^{\bol r}}$.
Within the ICMF approach, the finite-temperature dynamical susceptibilities
of ${\cal O}_A$ ($A=$SDW or SN) above 3D ordering temperatures are
calculated as
\begin{eqnarray}
\label{eq:Chi_RPA}
\chi_{A}(k_x,{\bol k},\omega) &=& \frac{\chi_A^{1D}(k_x,\omega)}
{1+J_{\rm eff}^A({\bol k})\chi_A^{1D}(k_x,\omega)},
\end{eqnarray}
where ${\bol k}=(k_y,k_z)$ is the wave vector in the
$y$-$z$ plane,
$\omega$ is the frequency, and
the effective coupling constants $J_{\rm eff}^A$ are given by
\begin{subequations}
\label{eq:J_eff}
\begin{align}
\label{eq:J_sdw}
J_{\rm eff}^{\rm SDW} ({\bol k}) =&
G_{\rm SDW}\sum_{\alpha=y,z}\big[J_{\alpha_1}\cos k_\alpha
-J_{\alpha_2}\sin(k_\alpha-\pi M)
\nonumber\\
&+J_{\alpha_3}\sin(k_\alpha+\pi M)
\big],
\\
J_{\rm eff}^{\rm SN} ({\bol k}) =& G_{\rm SN}\sum_{\alpha=y,z}
[{J_{\alpha_1}}^2
-(J_{\alpha_2}-J_{\alpha_3})^2]
\cos k_\alpha .
\label{eq:J_Ne}
\end{align}
\end{subequations}
The 1D susceptibilities
$\chi_A^{1D}(k_x,\omega)= 
\frac{1}{2}\sum_j e^{-ik_x j}
\int_0^\beta d \tau e^{i\omega_n\tau}\langle {\cal O}_A(j,\tau)
{\cal O}_A^\dag(0,0)\rangle |_{i\omega_n\to\omega+i\epsilon}$
are analytically computed by using the field theoretical technique 
($\beta=1/T$ and $\epsilon\to +0$)~\cite{Giamarchi_text}.
Those for SDW and spin-nematic operators respectively
take the maximum at $k_x^{\rm max}=(\frac12-M) \pi$ and $\pi$;
$\chi_{\rm SDW}^{1D}(k_x^{\rm max} ,0)=\frac{2}{v_+}
(\frac{4\pi }{\beta v_+})^{K_+/2-2}\sin(\frac{\pi K_+}{4})
B(\frac{K_+}{8},1-\frac{K_+}{4})^2$
and $\chi_{\rm SN}^{1D}(\pi,0)=\frac{2}{v_+}
(\frac{4\pi }{\beta v_+})^{2/K_+-2}\sin(\frac{\pi}{K_+})
B(\frac{1}{2K_+},1-\frac{1}{K_+})^2$,
where $B(x,y)$ is the beta function.

The transition temperature of each order is 
obtained from the divergent point of its susceptibility
at $\omega\to0$, which is given by
\begin{eqnarray}
\label{eq:transition}
1+{\rm Min}_{\bol k}[J_{\rm eff}^{A}({\bol k})]
\chi_{A}^{1D}(k_x^{\rm max},0) &=& 0.
\end{eqnarray}
The 3D ordered phase with the highest
transition temperature 
is realized. 
From this ICMF scheme, we can determine the phase diagram for
${\cal H}_{3D}$ with an arbitrary combination of 
$J_{y_i,z_i}$.
This is a significant advantage compared with previous theories for
spin-nematic phases.
We note that, when $J_{\rm eff}^A$ approaches
to zero, the present framework 
becomes less reliable and we need to consider subleading terms
in ${\cal H}_{\rm eff}^{3D}$.

From Eqs.~(\ref{eq:J_eff}) and (\ref{eq:transition}), we find that
the ordering wave numbers $k_{y,z}$ tend to be a commensurate value
$k_{y,z}=0$ or $\pi$ (see also Ref.~\onlinecite{note2}).
Thus the SDW ordered phase has the wave
vector $k_x=(\frac12-M)\pi$ and $k_{y,z}=0$ or $\pi$.
This agrees with the experimental result in
the intermediate-field phase of $\rm LiCuVO_4$~\cite{Masuda_VO4,Mourigal}.
For the spin-nematic ordered phase, we find the commensurate ordering vector
$(k_x,k_{y(z)})=(\pi,0)$ for $|J_{y_1(z_1)}|>|J_{y_2(z_2)}-J_{y_3(z_3)}|$
and $(k_x,k_{y(z)})=(\pi,\pi)$ for
$|J_{y_1(z_1)}|<|J_{y_2(z_2)}-J_{y_3(z_3)}|$.
This commensurate nature of $k_{x,y,z}$ in the nematic phase
is consistent with Ref.~\onlinecite{Zhitomirsky}.

\begin{figure}
\begin{center}
\includegraphics[width=8.5cm]{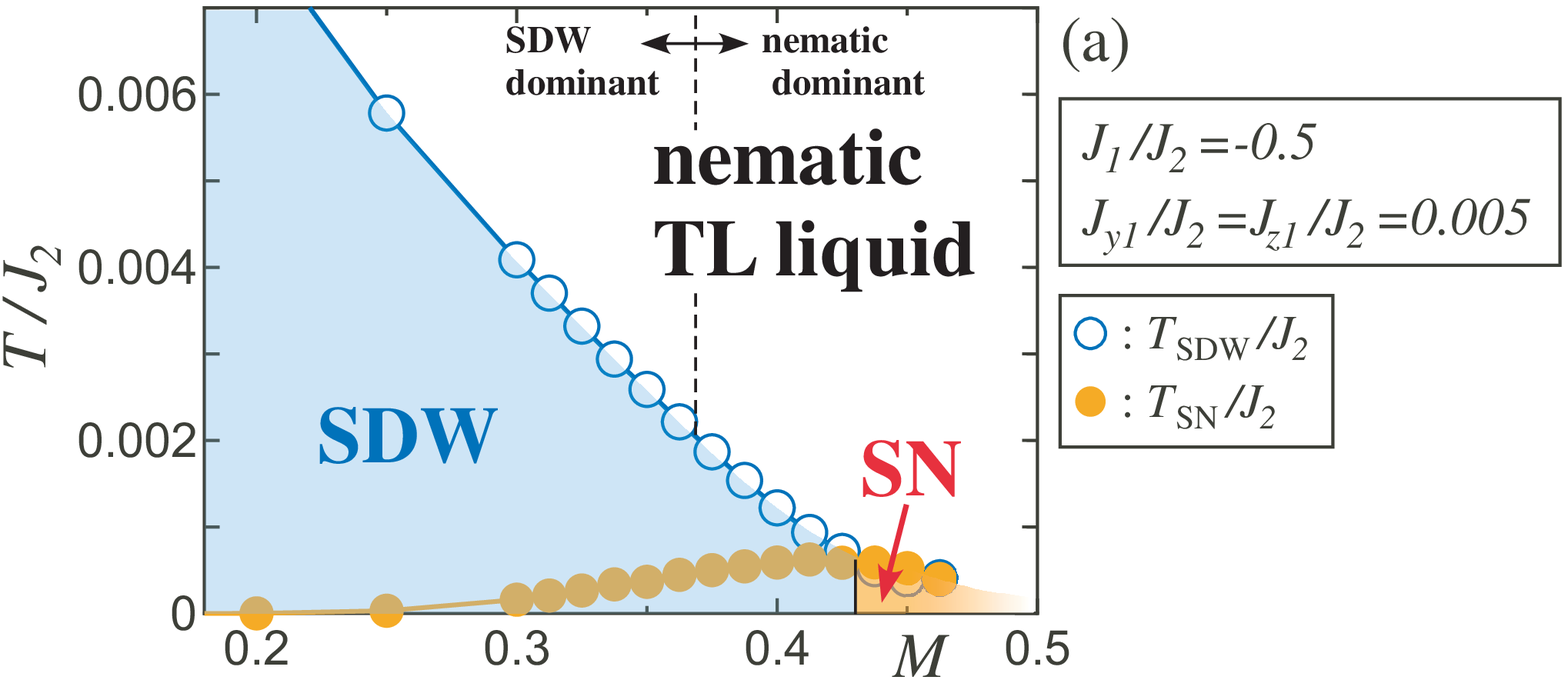}
\includegraphics[width=8.5cm]{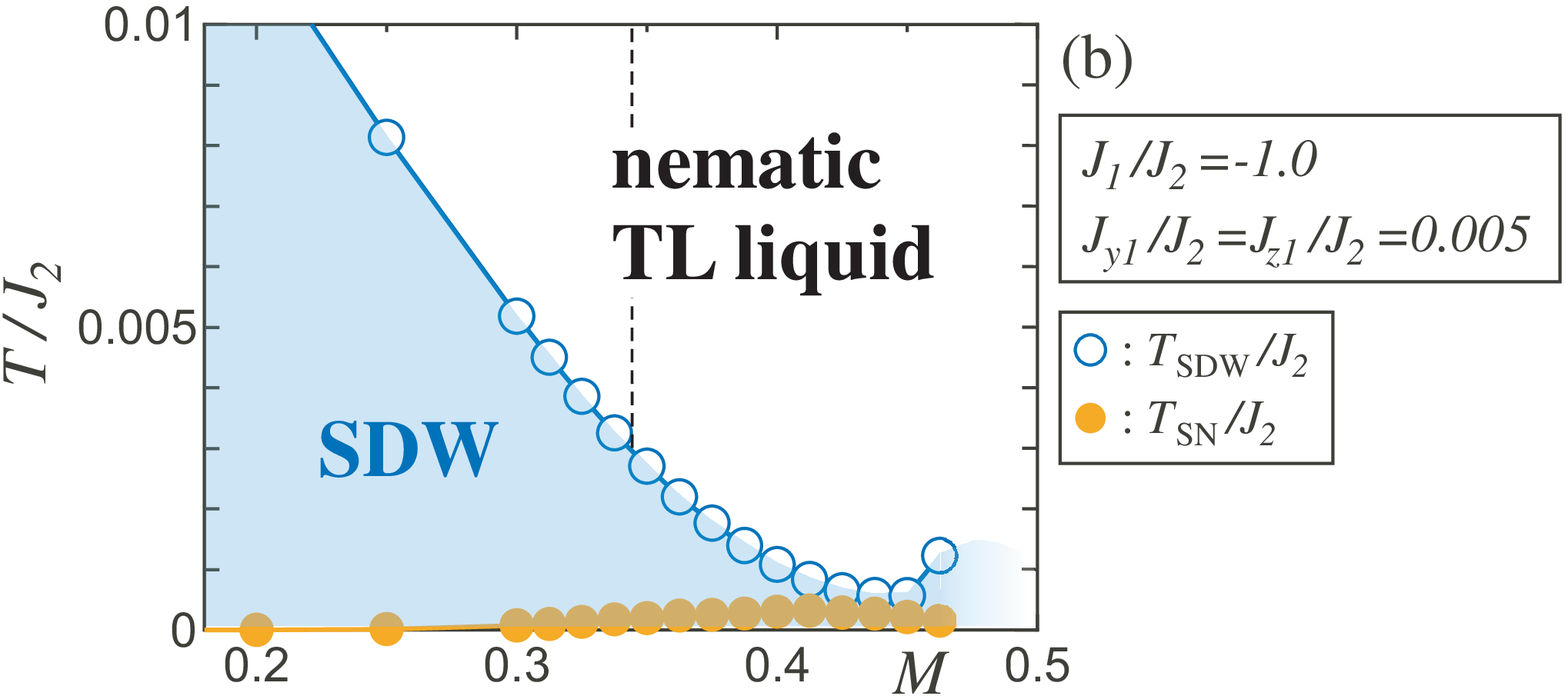}
\includegraphics[width=8.5cm]{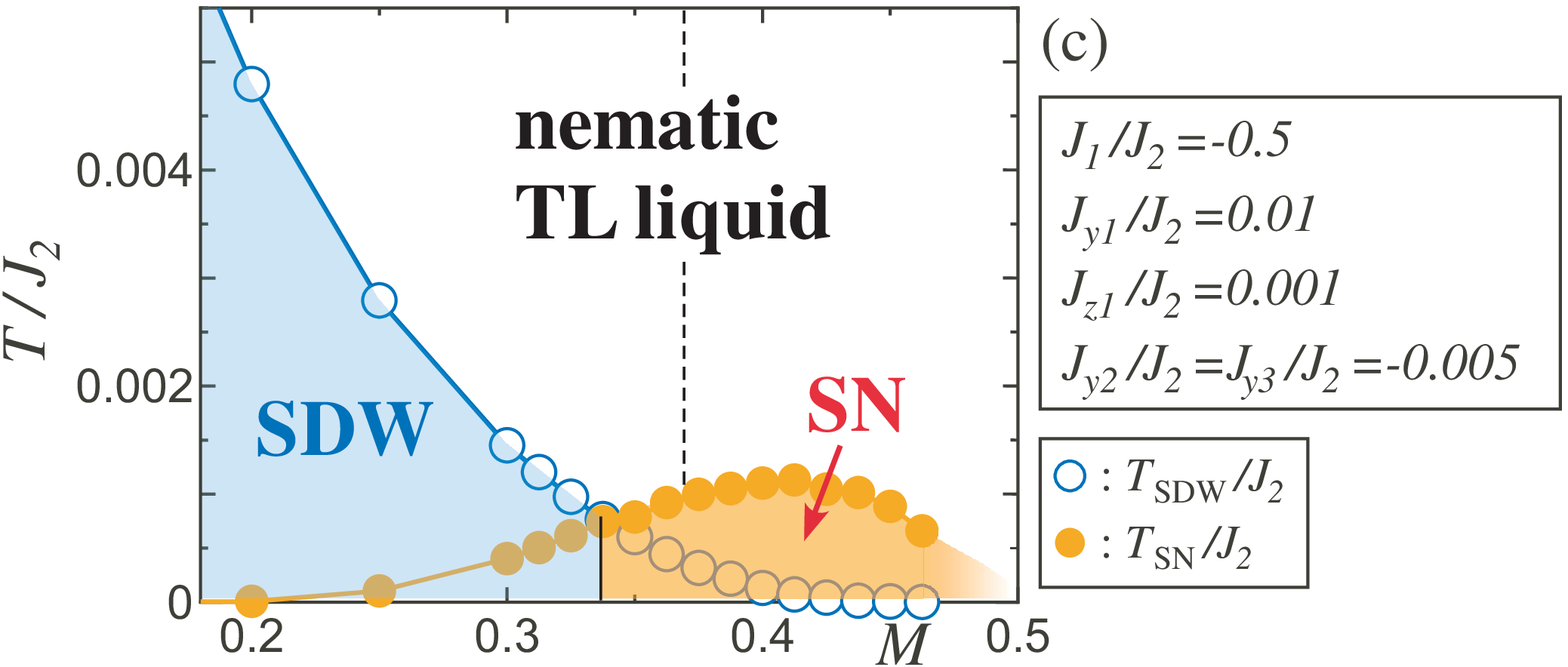}
\end{center}
\caption{(color online) Phase diagrams of the
weakly coupled $J_1$-$J_2$ chains (\ref{eq:Model})
in the $M$-$T$ plane, which are derived from the ICMF approach.
The temperatures $T_{\rm SDW(SN)}$ denote the
3D SDW (nematic) transition points.
The vertical dashed lines denote the crossover lines
between nematic dominant and SDW dominant TL liquids
in the 1D $J_1$-$J_2$ chain.
}
\label{fig:phasediagram}
\end{figure}

We show typical examples of obtained phase diagrams in
Fig.~\ref{fig:phasediagram}.
When interchain couplings are notfrustrated as the $J_{y_1,z_1}$
dominant cases of Figs.~\ref{fig:phasediagram}(a) and 
\ref{fig:phasediagram}(b), the SDW ordered phase is largely enhanced 
and the nematic ordered phase is
reduced to a higher-field regime compared to the crossover line
($K_+=2$) in the $J_1$-$J_2$ chain.
This is because the effective couplings $J_{\rm eff}^{\rm SDW}$ and
$J_{\rm eff}^{\rm SN}$ are respectively generated from the first- and
second-order cumulants, and therefore $J_{\rm eff}^{\rm SDW}$ is
generally larger than $J_{\rm eff}^{\rm SN}$ in non-frustrated systems with 
weak interchain couplings.
When both the couplings $J_{y_2,y_3}$ are dominant,
we find a similar tendency. 
We note that a model with dominant $J_{y_2,y_3}$ has been proposed 
for $\rm LiCuVO_4$~\cite{Enderle}, where 
a new phase expected to be a 3D nematic phase 
has been observed only near the saturation~\cite{Hagiwara}. 
From the calculations for the cases of $|J_1|/J_2=0.5$, 1.0, and 2.0,
we find that the nematic phase region in the $M$-$T$ phase diagram
generally becomes smaller with increase in $|J_1|/J_2$
since the value $g_-(x)$ in $G_{\rm SN}$ decreases.
When there is a certain frustration in interchain couplings, however,
the nematic phase region can expand, as shown in
Fig.~\ref{fig:phasediagram}(c).
When the signs of $J_{y_1}$ and $J_{y_2}(J_{y_3})$
are opposite, $J_{\rm eff}^{\rm SDW}$ becomes small, and 
the 3D nematic phase expands down to a relatively lower-field regime.
We emphasize that our theory succeeds in quantitatively
analyzing the competition between SDW and nematic ordered phases
in quasi-1D magnets.

\begin{figure}
\begin{center}
\includegraphics[width=8.5cm]{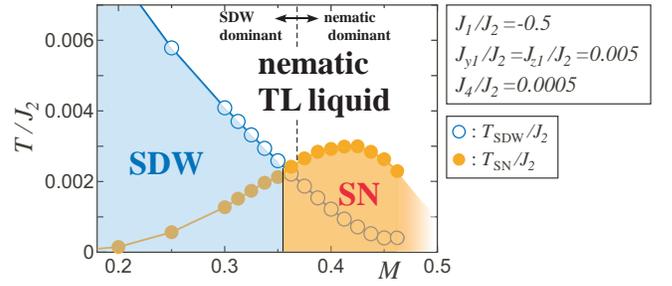}
\end{center}
\caption{(color online) Phase diagram of the weakly coupled
$J_1$-$J_2$ spin chains (\ref{eq:Model}) with 
a four-spin interaction ${\cal H}_4$.}
\label{fig:phasediagram2}
\end{figure}

\textit{Effects of four-spin term}.$-$
Finally, we study effects of an interchain four-spin interaction. 
The Hamiltonian we consider is
\begin{eqnarray}
\label{eq:4spin}
{\cal H}_4 &=& -J_4\sum_{j,\langle{\bol r},{\bol r}'\rangle}
S_{j,{\bol r}}^+S_{j+1,{\bol r}}^+S_{j,{\bol r}'}^-S_{j+1,{\bol r}'}^-
+{\rm H.c.}.
\end{eqnarray}
This interaction 
is a part of the spin-phonon coupling ${\cal H}_{\rm sp}=
-J_{\rm sp}\sum_{j,\langle{\bol r},{\bol r}'\rangle}
({\bol S}_{j,{\bol r}}\cdot{\bol S}_{j,{\bol r}'})
({\bol S}_{j+1,{\bol r}}\cdot{\bol S}_{j+1,{\bol r}'})$ 
and therefore it really exists 
in some compounds. One easily finds that
Eq.~(\ref{eq:4spin}) enhances the spin-nematic ordering. 
Applying the field theoretical strategy to 
the system ${\cal H}_{3D}+{\cal H}_4$,
we find that $J_{\rm eff}^{\rm SN}$ is
replaced with $J_{\rm eff}^{\rm SN}-4J_4C_0(\cos k_y+\cos k_z)$.
We thus obtain the phase diagram
for ${\cal H}_{3D}+{\cal H}_4$, as shown in Fig.~\ref{fig:phasediagram2}.
Comparing Figs.~\ref{fig:phasediagram}(a) and \ref{fig:phasediagram2}, 
we see that an inter-chain
four-spin interaction definitely enhances the 3D nematic phase
even if its coupling constant $J_4$ is small. Since $J_4$ is usually
positive, it favors ferrotype nematic ordering along the $y$ and $z$ 
axes; i.e., $k_{y,z}=0$.

\textit{Conclusion}.$-$ We have constructed
finite-temperature phase diagrams for 3D spatially anisotropic magnets,
which consist of weakly coupled spin-$\frac{1}{2}$ $J_1$-$J_2$ chains,
in an applied magnetic field. Incommensurate SDW and spin-nematic ordered
phases appear at sufficiently low temperatures, triggered by the 
nematic TL-liquid properties in the $J_1$-$J_2$ spin chains. We reveal
several natures of orderings in the coupled $J_1$-$J_2$ chains:
The 3D nematic ordered phase is generally smaller than the 1D nematic dominant
region, while it can be larger if we somewhat tune the inter-chain
couplings. The ordering wave numbers $k_{y,z}$ tend to be $0$ or $\pi$,
and a small four-spin interaction ${\cal H}_4$ efficiently helps
the 3D nematic ordering. We finally note that our theory can also be 
applied to AF-$J_1$ systems.

\textit{Acknowledgements}.$-$
We thank Akira Furusaki for fruitful discussions at the early stage
of this study.
This work was supported by KAKENHI No.\ 21740295, No.\ 22014016, and
No.\ 23540397 from MEXT, Japan.


\begin{thebibliography}{99}
\bibitem{note1}
One can easily take account of other different inter-chain
couplings using our theoretical framework.
\bibitem{note2}
Correctly speaking, $G_{\rm SDW}$ can take both positive
and negative values
$\pm A_1^2|\langle e^{i\sqrt{\pi}\phi_-}\rangle_-|^2$
due to degenerate pinning positions
$\sqrt{4\pi}\phi_- + \pi M = (2n+\frac{1}{2})\pi$ ($n$: integer)
in each $J_1$-$J_2$ chain. This ambiguity is lifted
by subleading inter-chain terms omitted in ${\cal H}_{\rm eff}^{3d}$.
We note that the 3D SDW ordering temperature does not depend on
the sign of $G_{\rm SDW}$ in the weak-coupling regime
$|J_{y_i,z_i}|\ll |J_{1,2}|$.
\end{thebibliography}
\end{document}